%% file: main.tex
\documentclass{webofc}
\usepackage[varg]{txfonts}   
\usepackage{xcolor}

\usepackage{subcaption}
\usepackage{listings}
\usepackage{makecell}
\usepackage{natbib}
\input{macros}

\begin{document}
\title{Apprentice for Event Generator Tuning}
%
%

\author{\firstname{Mohan} \lastname{Krishnamoorthy}\inst{1}\fnsep\thanks{\email{mkrishnamoorthy@anl.gov}} \and
        \firstname{Holger} \lastname{Schulz}\inst{2}\fnsep\thanks{\email{holger.schulz@durham.ac.uk}} \and
        \firstname{Xiangyang} \lastname{Ju}\inst{4}
        \and
        \firstname{Wenjing} \lastname{Wang}\inst{4}
        \and
        \firstname{Sven} \lastname{Leyffer}\inst{1}
        \and
        \firstname{Zachary} \lastname{Marshall}\inst{4}
        \and
        \firstname{Stephen}
        \lastname{Mrenna}\inst{3}
        \and
        \firstname{Juliane} \lastname{M\"uller}\inst{4}        
        \and
        \firstname{James B.} \lastname{Kowalkowski}\inst{3}
}

\institute{Argonne National Laboratory, Lemont, IL 60439
\and
           Department of Computer Science, Durham University, South Road, Durham DH1 3LE, UK 
\and
Fermi National Accelerator Laboratory, Batavia, IL 60510
\and
Lawrence Berkeley National Laboratory, Berkeley, CA 94720
 }

\abstract{%
  \apprentice is a tool developed for event generator tuning.
  It contains a range of conceptual improvements and extensions over the tuning tool \professor.  Its core functionality remains the construction of a multivariate analytic surrogate model to computationally expensive Monte-Carlo event generator predictions.  The surrogate model is used for numerical optimization in chi-square minimization and likelihood evaluation.   
{\it Apprentice} also introduces algorithms to automate the selection of observable weights to minimize the effect of mis-modeling in the event generators.  We illustrate our improvements for the task of MC-generator tuning and limit setting.
}
\maketitle

\input{introduction}
\input{apprenticeProblems}
\input{apprenticeUse}

\input{apprenticeImpact}
\input{conclusion}

\section{Acknowledgments}
 
This work was supported by the U.S. Department of Energy, Office of Science, Advanced Scientific Computing Research, under Contract DE-AC02-06CH11357.
Support for this work was provided through the Scientific Discovery through Advanced Computing (SciDAC) program funded by U.S. Department of Energy, Office of Science, Advanced Scientific Computing Research, grant ``HEP Data Analytics on HPC'', No.~1013935.
This work was also supported by
the U.S. Department of Energy through grant DE-FG02-05ER25694, and
by Fermi Research Alliance, LLC under Contract No. DE-AC02-07CH11359 with the U.S. Department of Energy, Office of Science, Office of High Energy Physics.
This work was supported in part by the U.S. Department of Energy, Office of Science, Office of Advanced Scientific Computing Research and Office of Nuclear Physics, SciDAC program through the FASTMath Institute under Contract No. DE-AC02-05CH11231 at Lawrence Berkeley National Laboratory.

\bibliography{NLP,robust,minlp,jshort,ndrathep,ra,hep}

%
%
%
\vfill
\begin{flushright}
	\scriptsize
	
	\framebox{\parbox{\textwidth}{
			The submitted manuscript has been created by UChicago Argonne, LLC, Operator of Argonne National Laboratory (“Argonne”). 
			Argonne, a U.S. Department of Energy Office of Science laboratory, is operated under Contract No. DE-AC02-06CH11357. 
			The U.S. Government retains for itself, and others acting on its behalf, a paid-up nonexclusive, irrevocable worldwide 
			license in said article to reproduce, prepare derivative works, distribute copies to the public, and perform publicly 
			and display publicly, by or on behalf of the Government.  The Department of Energy will provide public access to these 
			results of federally sponsored research in accordance with the DOE Public Access Plan. 
			\url{http://energy.gov/downloads/doe-public-access-plan}.
	}}
	\normalsize
\end{flushright}

\end{document}

%% file: macros.tex
\usepackage{xspace,amsmath,booktabs,graphicx}
\usepackage{lipsum}
\usepackage[T1]{fontenc} 

\newcommand{\MC}{\ensuremath{\text{MC}}\xspace}

\newcommand{\chisq}{\ensuremath{\chi^2}\xspace}

\renewcommand{\vec}[1]{\ensuremath{\mathbf{#1}}\xspace}
\newcommand{\p}{\vec{p}}
\newcommand{\w}{\vec{w}}

\newcommand{\professor}{\textsc{Professor}\xspace}
\newcommand{\apprentice}{\textsc{apprentice}\xspace}

\newcommand{\MeV}{\text{Me\kern -0.15ex V}\xspace}
\newcommand{\GeV}{\text{Ge\kern -0.15ex V}\xspace}
\newcommand{\TeV}{\text{Te\kern -0.15ex V}\xspace}

\newcommand{\calO}{\mathcal{O}}

\newcommand{\calR}{\mathcal{R}}
\newcommand{\calS}{\mathcal{S}}

%% file: introduction.tex
\section{Introduction}
\label{intro}

Monte Carlo-based (MC) event generators are necessary tools for interpreting data at the high energy frontier.
MCs contain {\cal O}(10-100) parameters that are tuned to match
selected data to allow predictions for other sets of data.   In many cases, the limiting factor
to precision physics at the LHC is our lack of confidence in the MC predictions.  The tuning task is daunting because
of the large number of parameters that should be explored and the computational cost of simulations.
For the large datasets of collider data that are available for MC tuning, a common heuristic to evaluate
the goodness of a prediction is:
\begin{equation}
\chisq(\p,\w)= \sum_{\calO\in\calS_\calO}w_\calO \sum_{b\in\calO}\frac{(t_b(\p)-\calR_b)^2}{\Delta t_b(\p)^2+\Delta\calR_b^2}, \label{eq:poly}
\end{equation}
where $\calS_\calO$ is the set of observables $\calO$ used in the tune, 
each observable has a weight $w_\calO$ represented by a vector $\w$,
$t_b(\p)/\calR_b$ is the theory prediction/reference data in a given bin $b$ of an observable and
the $\Delta t_b/\Delta \calR_b$'s are error estimates on these quantities.

Our problem is to minimize $\chisq(\p,\w)$ as a function of the
adjustable parameters \p and possibly the observable weights \w.
To accomplish this, one needs a range of theory predictions $t_b$ for
different possible parameter choices \p and a method or principle for choosing \w.
As an additional output, it is desirable to have an estimate of what range of parameters \p are compatible with
the data at a given confidence level.

In the following, we report on \apprentice, a successor to the tool \professor \cite{professor} that was developed to
accomplish these goals.

%% file: apprenticeProblems.tex
\section{Problems Solved using \apprentice}
\label{sec:appprob}

\professor is a public tool that was introduced to automate and accelerate the tuning of event generators.
Based on a modest number of MC simulations, it constructs a polynomial approximation surrogate to these predictions in each
bin of a histogram representing the observables that drive the tuning.
Given the surrogate function, an "optimal" set of tuned parameters \p is determined by minimizing the
heuristic (\ref{eq:poly}).

\professor has been used successfully to create a number of MC tunes and in other contexts.
Nonetheless, there are certain aspects that could be improved.
First, the polynomial fit to the MC predictions is adequate for well-behaved distributions, but does not perform
well for a prediction $x$ with distribution $\sim 1/x$.   Second, the suite of minimizers available
is limited to those in SciPy and the CERN Minuit package.    The dependence on minimizer, if any, needs to
be explored.   Third, the choice of weights $w_\calO$ is done manually, and any optimization over them must
be done in a costly, iterative process.    An algorithm for automatically selecting these weights is desirable.

The desire to address these three issues has led to the development of a new public tool that is sufficiently different
from its predecessor to have its own name: \apprentice.

In the following, we motivate the use of \apprentice by listing the problems that \apprentice can currently solve and by highlighting the advantages of using \apprentice to solve these problems, especially for HEP applications. 

\subsection{Rational Approximation as a Surrogate Function}
\label{sec:ratApprox}

Polynomial models are relatively easy to build and use.
However, they have poor extrapolation behavior and
are severely limited in their ability to cope with singularities.  These
drawbacks can reduce their effectiveness at representing physics models.
On the other hand, rational functions (quotients of polynomials) can
be considerably more effective \citep{Devore1986,New1964} at 
representing models that have real or apparent pole structure.
Unfortunately, rational approximations can be numerically fragile to compute
and are prone to having spurious singularities.  

\apprentice is capable of computing 
multivariate rational approximations $r(x)=p(x)/q(x)$ to simulation data using
three approaches. The first is
based on the univariate methods of \citep{GPT11,PGV12} and provides a robust and
efficient way to compute the coefficients of $p(x)$ and $q(x)$.  Although it
tries to reduce the appearance of unwanted
singularities by using ideas from linear algebra to minimize the degree of
$q(x)$, it does not \emph{guarantee} that $r(x)$ will be pole free in the
parameter domain $D$.

The second uses a constrained optimization formulation that includes
structural constraints on $r(x)$ to enforce the absence of poles in $D$.
Although this approach is computationally more expensive than the
first, it guarantees that the computed
approximation is free of poles for box-shaped parameter domains, 
which can be crucial when computing surrogate models for use in optimization.
In particular, the guaranteed absence of poles ensures that
subsequent optimization problems involving our rational approximations are
well-defined. 

The third approach is a specialization of the second that is used to efficiently find a pole-free
rational approximation by making $q(x) = \mathbf{A}^T\mathbf{x}+\mathbf{b}$ linear.
This is achieved using auxiliary variables to write the dual of the linear program $\min_\mathbf{x}~\mathbf{A}^T~\mathbf{x}~+~\mathbf{b}~s.t.~\mathbf{x}~\in~[\mathbf{L},\mathbf{U}]$ as a constraint to force the denominator to be greater than 0 or be pole-free. 
The advantage of this approach is that it requires solving the optimization problem only to find a pole-free rational approximation (as opposed to an iterative approach) and hence it is more efficient.

\subsection{Tuning Problem}
\label{sec:tuningprob}
The goal of the tuning problem is to find an optimal set of physics parameters, $\p^*$, that  minimizes the
function (\ref{eq:poly}).
When the theory prediction $t_b(\p)$ is based on simulated data from the MC event generator, the computational
cost is high (the generation of 1 million events for a given set of parameters consumes about 800 CPU minutes on a typical computing cluster), severely
limiting the number of parameter choices $\p$ that can be explored in the tuning.   
To overcome this issue, we construct a parametrization of a modest number of  MC simulations i.e., of $\MC_b(\p)$ and $\Delta \MC_b(\p)$
of each bin b as a polynomial or rational approximation using \apprentice as discussed in the previous section.
Thus $t_b(\p)$ becomes our analytic surrogate $f_b(\p)$, which is easy to compute (it can be evaluated in
milliseconds) and minimize.
For performing this optimization, the solver options of Truncated Newton method~\cite{nocedal2006numerical}, Newton-conjugate gradient method~\cite{shewchuk1994introduction}, LBFGS-B algorithm~\cite{byrd1995limited}, and Trust region algorithm~\cite{conn2000trust} from python's SciPy package are provided within \apprentice.

\subsection{Automatic Selection of Observable Weights}
\label{sec:tuningautoselection}
To tune parameters as discussed in the previous section, the weights $w_\calO$ in $\chisq(\p,\w)$ need to be specified. In practice, the weights are adjusted manually, based on experience and physics intuition: the expert fixes the weights and minimizes (\ref{eq:poly}) over the parameters $\p$.
If the resulting fit is unsatisfactory, a new set of weights is selected, and the optimization over $\p$ is repeated until the tuner is satisfied.
Our  goal is to automate the weight adjustment, yielding a less subjective and less time-consuming process.

In \apprentice, the automatic selection of weights is achieved using two mathematical formulations: bilevel and robust optimization.
In bilevel optimization, a merit function is minimized over the set of weights.
\apprentice provides three options for merit functions.
The first is the portfolio objective function, which is motivated by portfolio optimization in finance, where the  goal is to maximize the expected return while  minimizing  the risk.   Translated to our problem, we want to minimize the expected error over all observables while also minimizing the variance over these errors. The second and the third merit functions are based on the central values for scoring schemes with a scoring rule, $S(P,x) = -\left(\frac{x-\mu_P}{\sigma_P}\right)^2 - \log \sigma_P^2$,
where model $P$ has mean performance $\mu_P$ and variance $\sigma_P^2$.
For our application, $x$ corresponds to the simulation prediction $f_b(\p)$,  $\mu_P$ to our observation data $\calR_b$,  and the variance $\sigma^2_P$
to our data uncertainty $\Delta\calR_b$.
The second merit function maximizes the sum over all observables of the mean across all bins in that observable of each bin's scoring function.
The third merit function maximizes over the median.  The scoring function for bin b is defined as $\displaystyle \Bigg\{\left(\frac{f_b(\widehat{\p}_{\w})-\calR_b}{\Delta\calR_b}\right)^2 + \log(\Delta\calR_b^2)\Bigg\}$.
Since we don't have the full analytical expression of the outer objective function, this function is approximated with a radial basis function (RBF) \cite{Powell1999} to perform the bilevel optimization in \apprentice.

Robust optimization is a single-level formulation for finding the optimal weights for $\chisq(\p, \w)$.
It estimates the parameters ${\p}$ that minimize
the largest deviation $\left(f_b(\p) - \mathcal{R}_b\right)^2$ over
all bins in an uncertainty set $\mathcal{U}_b$ of bin $b$.
Assuming that the experiment and the MC simulation are described using
independent random variables with mean $\mathcal{R}_b$, the
uncertainty set $\mathcal{U}_b$ for each bin $b$ is described by the
interval $[\mathcal{R}_b-\Delta\calR_b-\Delta
f_b(\p),\mathcal{R}_b+\Delta\calR_b +\Delta f_b(\p)]$.
Since $\left(f_b(\p) - \mathcal{R}_b\right)^2$ is a square function, the largest deviation occurs at either end of the interval. Using this,  the robust optimization formulation can be easily rewritten as a one shot single level optimization problem.
To avoid the trivial solution of all weights being zero,
the sum of the weights $\w\in[0,1]$ across all the observables is required to cover at least $\mu$\% of the total observables in $\calS_\calO$, where $\mu$ is a hyperparameter set by the end user.

%% file: apprenticeUse.tex
\section{Using \apprentice}
\label{sec:appuse}

The source code for \apprentice is available at \url{https://github.com/HEPonHPC/apprentice}. For each problem described in Section~\ref{sec:appprob}, we provide a convenient script that the end user can use directly with appropriate input arguments to solve that problem for their use case. In this section, we first describe how to clone and install \apprentice, and, then, how the interface scripts can be used by the end user to directly solve their problem. For this, we describe the important arguments of each script in this section. For the complete usage, run the script using the \verb|-h| flag. Then, we describe what output to expect from each of these scripts. Additionally, we also note the software features that are currently implemented for each problem. 
\subsection{Setting up \apprentice}
To clone and install apprentice, the following commands can be used from a Unix Terminal:

\begin{lstlisting}[language=bash]
$git clone --recurse-submodules git@github.com:HEPonHPC/apprentice.git
$cd apprentice
$pip install .
$cd pyoo
$pip install .
\end{lstlisting}

\subsection{Creating Polynomial and Rational Approximations}
\label{sec:createapprox}
We provide an easy-to-use script at \verb|apprentice/bin/app-build| to create a rational approximation $r(x)=p(x)/q(x)$ for an arbitrary degree for $p(x)$ and $q(x)$ or when the order of $q(x)$ is one or zero i.e., polynomial approximation. 
The important arguments of the script are:
(a) \textit{args[0]}: Location of directory of YODA files or of the HDF5 file with the MC generator central values and uncertainty values for all bins to build approximations for, 
(b) \textit{--order}: order of numerator polynomial and denominator polynomial separated by a comma e.g., \textit{2,1}, 
(c) \textit{--errs}: Boolean, which if True then the approximation $\Delta f_b(\p)$ is created for MC generator uncertainty values, otherwise the approximation $f_b(\p)$  is created for MC generator central values,
(d) \textit{--pnames}: data file with parameter names (one name per line of the text file), 
(e) \textit{-t}: Number of multistarts to use for pole detection, and 
(f) \textit{-o}: Output JSON file location.

\footnotetext[1]{The order of the coefficients can be found in~\url{people.sc.fsu.edu/~jburkardt/py_src/monomial/monomial.html}}
The output is a JSON object, where the keys at the first level are the bin names, and, at the second level, include:
(a) \textit{dim}: Number of parameter dimensions,
(b) \textit{m}: order of numerator polynomial, 
(c) \textit{n}: order of denominator polynomial (present if order of denominator polynomial is greater than 0),
(e) \textit{pcoeff}: numerator coefficients\footnotemark, and
(f) \textit{qcoeff}: denominator coefficients (present if order of denominator polynomial is greater than 0).\footnotemark[\value{footnote}]

\subsection{Solving the $\chi^2$ Tuning Problem}
\label{sec:chi2solutionuse}
For solving the tuning problem described in Section~\ref{sec:tuningprob}, the interface script is located at \verb|apprentice/bin/app-tune2|. 
The important arguments of the script are:
(a) \textit{args[0]}: Weights text file location with two columns (separated by a space) where the first column are the observable names $\calO\in\calS_\calO$ and the second column contain $w_\calO$.
(b) \textit{args[1]}: Measurement data central and uncertainty values in JSON format where the first level of keys are the bin names and the corresponding value is a two-element array containing central and uncertainty values,
(c) \textit{args[2]}: Approximation JSON file location created using the script described in Section~\ref{sec:createapprox} for the MC generator central values,
(d) \textit{-e}: Approximation JSON file location created using the script described in Section~\ref{sec:createapprox} for the MC generator uncertainty values,
(e) \textit{-s|--survey}: Size of survey when determining the starting point,
(f) \textit{-r|--restart}: Number of restarts to use with different starting points with the aim of finding multiple local minima,
(g) \textit{--msp}: Manual start parameter vector to use (values separated by a comma), and
(h) \textit{-a|--algorithm}: The minimization algorithm, options include: \textit{tnc} for Truncated Newton method, \textit{ncg} for Newton-conjugate gradient method, \textit{lbfgsb} for LBFGS-B algorithm, and \textit{trust} for Trust region algorithm.

The output of this script are the parameters $\p^*, \chi^2(\p^*,\w)$, and some other pertinent information about the optimization for logging purposes. This script also supports running the optimization with multiple starting points in parallel using mpi4py~\cite{MPI4py}.

\subsection{Event Generator Tuning by Automatic Selection of Observable Weights}
In this section, we describe the interface scripts for solving generator tuning problems with automatic selection of observable weights. 
\subsubsection{Bilevel Optimization Formulation}
Solving the generator tuning problem with selection of observable weights using the bilevel optimization formulation as described in Section~\ref{sec:tuningautoselection} requires two steps: (a) generating the data to train the RBF that will be used as an approximation of the outer level objective function and (b) solving the bilevel optimization problem with either of the three merit functions of the portfolio function, the mean of the scoring function (meanscore), or the median of the scoring function (medianscore).
The interface script for generating the training data is at \verb|apprentice/pyoo/bin/pyoo-train|. 
The arguments \textit{args[0]}, \textit{args[1]}, \textit{args[2]}, \textit{-e}, \textit{-s|--survey}, and \textit{-r|--restart} of this script mean the same as those described in Section~\ref{sec:chi2solutionuse}.
Additional (important) arguments of the script are: 
(a) \textit{-d|--design} Size of the initial design,
(b) \textit{-o|--output} Training data output file, and
(c) \textit{--seed} Random seed to use.

Using the training output data (a JSON file), we can run the interface scripts for solving the bilevel optimization using the portfolio objective found at \verb|apprentice/pyoo/bin/pyoo-run-portfolio|, meanscore objective found at \verb|apprentice/pyoo/bin/pyoo-run-meanscore|, and medianscore objective found at \verb|apprentice/pyoo/bin/pyoo-run-medianscore|. The arguments for all three scripts are the same. 
Also, the arguments \textit{args[0]}, \textit{args[1]}, \textit{args[2]}, \textit{-e}, and \textit{-s|--survey} of these scripts mean the same as those described in Section~\ref{sec:chi2solutionuse}.
Additional (important) arguments of the script are: 
(a) \textit{args[3]}: Training data JSON file,
(b) \textit{-n|--niterations}: Number of iterations to use, and
(c) \textit{-o|--output}: Output JSON file.

The output of each script is a JSON object with keys such as \textit{X\_outer}, \textit{X\_inner}, \textit{Y\_outer}, \textit{hnames} and \textit{pnames}. 
The keys \textit{X\_outer} and \textit{X\_inner} point to an array of array where each array element in the outer array are the weights $\w^*$ of the observables and the parameters $\p^*$ obtained in each iteration of the outer optimization.
The best weight and parameter element in the output corresponds to the index of the smallest value in the array against the key \textit{Y\_outer} i.e., smallest outer objective value.
The order of the observables and parameter dimensions that the weights and parameter values correspond to can be found against the keys \textit{hnames} and \textit{pnames}, respectively.

\subsubsection{Single Level Robust Optimization Formulation}

To solve the generator tuning problem where the observable weights are selected using the robust optimization approach (see Section~\ref{sec:tuningautoselection}), the interface script is located at  \verb|apprentice/pyoo/bin/pyoo-robopt|.
The important arguments of the script are:
(a) \textit{-w|--weights}: Weight file (see description of \textit{arg[0]} in Section~\ref{sec:chi2solutionuse}),
(b) \textit{-d|--expdata}: Measurement data (see description of \textit{arg[1]} in Section~\ref{sec:chi2solutionuse}),
(c) \textit{-a|--approx}: MC central values approximation JSON file (see description of \textit{arg[2]} in Section~\ref{sec:chi2solutionuse}),
(d) \textit{-e|--error}: See description of \textit{-e|--error} in Section~\ref{sec:chi2solutionuse},
(e) \textit{-o|--outtdir}: Output directory where the output JSON file and the optimization log files are stored (more details below),
(f) \textit{-s|--solver}: Solver where the options are \textit{ipopt} that makes use to IPOPT solver with AMPL solver interface (ASL) using PYOMO~\cite{hart2017pyomo,hart2011pyomo} and {cyipopt}~\cite{cyipopt} uses the python wrapper around IPOPT and avoids making use of PYOMO and the ASL, and
(g) \textit{-m|--mu}: Hyperparameter $\mu \in [0,1]$ is the value to be used in the constraint to avoid tivial solutions of all weights being zero.

The output of the script is a JSON object where the weights $\w^*$ of the observables are in the array pointed by the key \textit{X\_outer}, the parameters $\p^*$, which is obtained by solving the $\chi^2$ tuning problem using $\w^*$, are in the array pointed by the key \textit{X\_inner}. The order of the observables and parameter dimensions that the weights and parameter values correspond to can be found against the keys \textit{hnames} and \textit{pnames}, respectively.
Additionally, the object pointed by the key \textit{log} contains some pertinent information about the optimization for logging purposes.




%% file: apprenticeImpact.tex
\section{Impact of \apprentice for HEP applications}
\label{sec:appimpact}

In this section, we discuss the impact of the problems that can be solved using \apprentice for HEP applications.

\subsection{Rational Approximation for High Energy Physics}\label{SEC:HEP}
As a demonstration of the utility of rational approximations, we
consider the problem of setting limits on dark matter particles in a
future direct detection Xenon-based experiment
(see \cite[Sec. 26]{PhysRevD.98.030001}).   
The  physics model has three parameters, $x=(m_\chi, c_+, c_\pi)$, representing
the dark matter particle mass (ranging over $10$-$100 \mathrm{GeV/c^2}$) and two
(dimensionless) couplings to ordinary matter (ranging from
$10^{-4}$-$10^{-3}$ and $10^{-3}$-$10^{-1}$, respectively).
We consider a ``binned likelihood'' analysis \citep{Barlow:1993dm} of 
$\displaystyle \mathcal{L}(x^{(k)}|d_1, d_2, \ldots, d_6) = \prod_{b=1}^6 \frac{N_b(x^{(k)}) d_b e^{N_b(x^{(k)})}}{\Gamma(d_b+1)},$
where the $d_b$ represent six data points and $N_b(x^{(k)})$ denote the simulated quantities for a point $x^{(k)}$
that correspond to the $d_b$.
Since no measured data exists as yet, 
 we assume a signal consistent with a dark matter mass $m_\chi=$ 10
 $\mathrm{GeV/c^2}$ and an interaction strength large enough to
 produce approximately 100 events, yielding
 $\{d_1, d_2, \ldots, d_6\} =\{ 70.4 , 26.7 , 9.8 , 3.4 , 1.0 , 0.2 \}$,
simulated by using a specific framework of the generalized
spin-independent response to dark matter in direct detection experiments~\citep{Hoferichter:2016nvd,Cerdeno:2018bty}. 
Regions of parameter space consistent with the simulated data are
found using MultiNest~\citep{Feroz:2008xx,Feroz:2013hea,pymultinest}.
This requires the evaluation of the likelihood function at tens of
thousands\footnote{The dimension of the problem and the convergence criteria of
	the MultiNest algorithm strongly influence the number of required function
	calls.} of $x^{(k)}$ to succeed, and  the computational cost
      can be  substantial.
To reduce the cost, we replace the expensive simulation of $N_b$ with the cheaper
evaluation of a rational approximation surrogate.
The results presented here are for rational approximations of degree $M=4,N=4$ as well as polynomial approximations
of degree 7.\footnote{The degree is chosen such that the number of coefficients is comparable to the number of coefficients used in the
	rational approximations.}
The maximization of the likelihood function requires about 30,000
evaluations in all cases, but is about a factor of 50 times faster
using the rational or polynomial approximation.

\begin{figure}[htb!]
	\centering
	\begin{subfigure}[b]{0.24\textwidth}
		\centering
		\includegraphics[width=.9\textwidth]{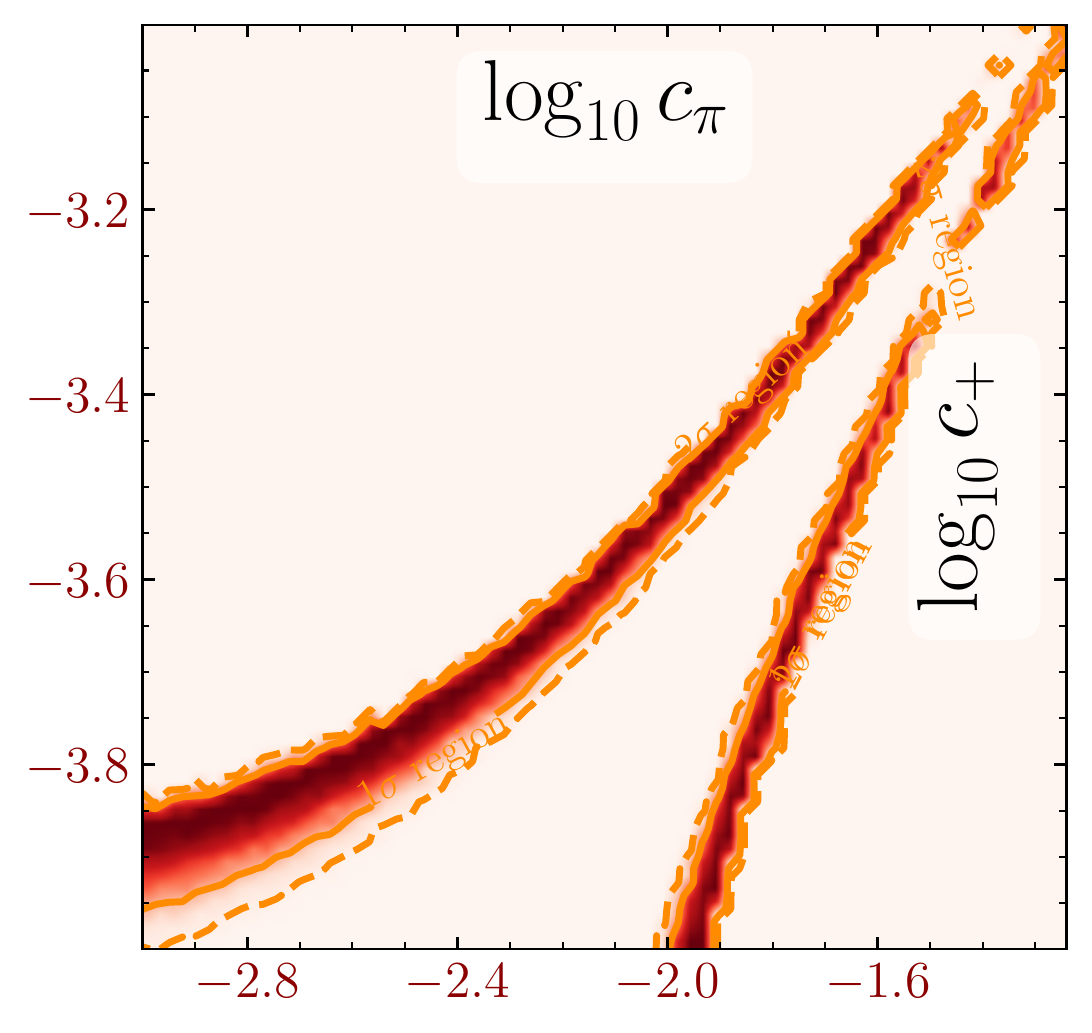}
		\caption%
		{{\small Full simulation using MC generator}}    
		\label{fig:nest-si-full}
	\end{subfigure}
	\hfill
	\begin{subfigure}[b]{0.24\textwidth}  
		\centering 
		\includegraphics[width=.9\textwidth]{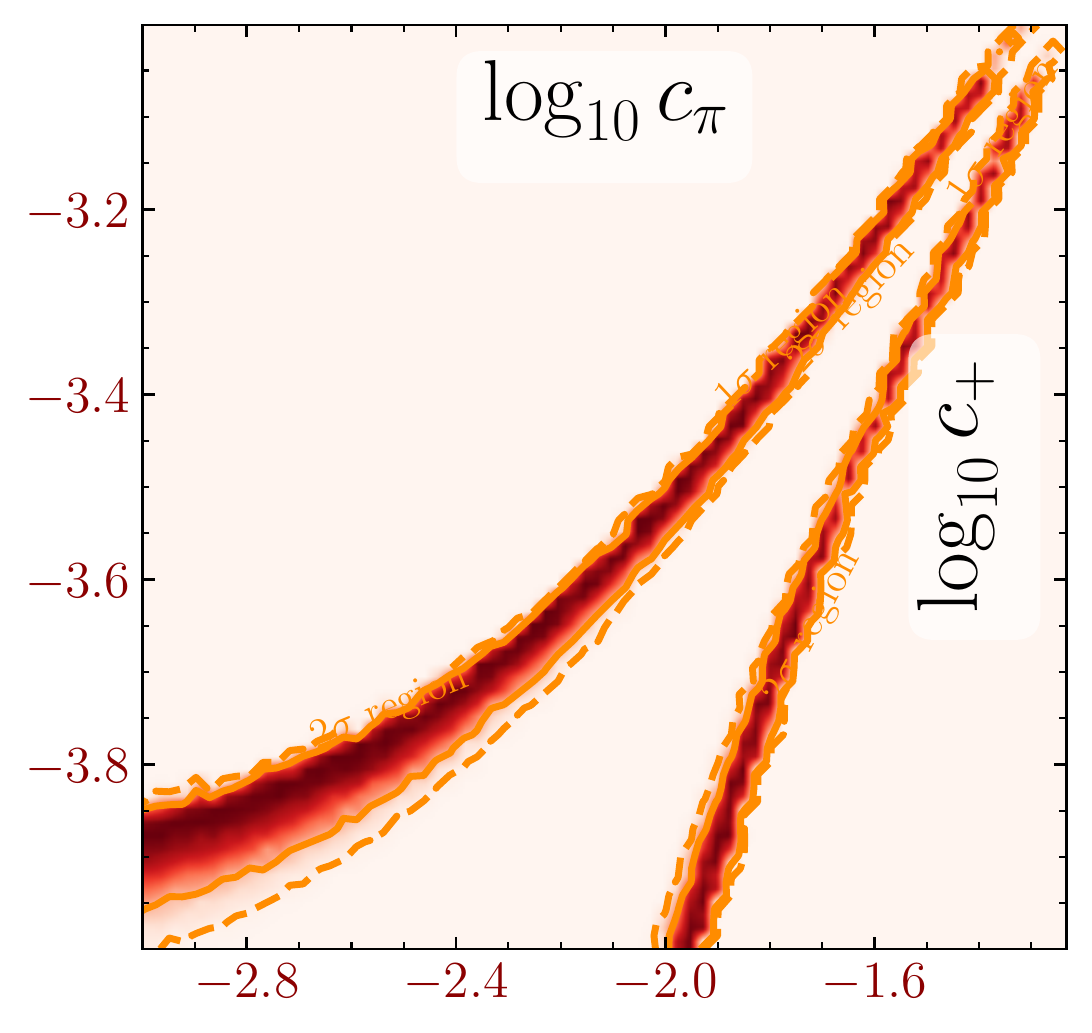}
		\caption[]%
		{{\small Pole-free rational approximation}}    
		\label{fig:nest-si-sip}
	\end{subfigure}
	\hfill
	\begin{subfigure}[b]{0.24\textwidth}   
		\centering 
		\includegraphics[width=.9\textwidth]{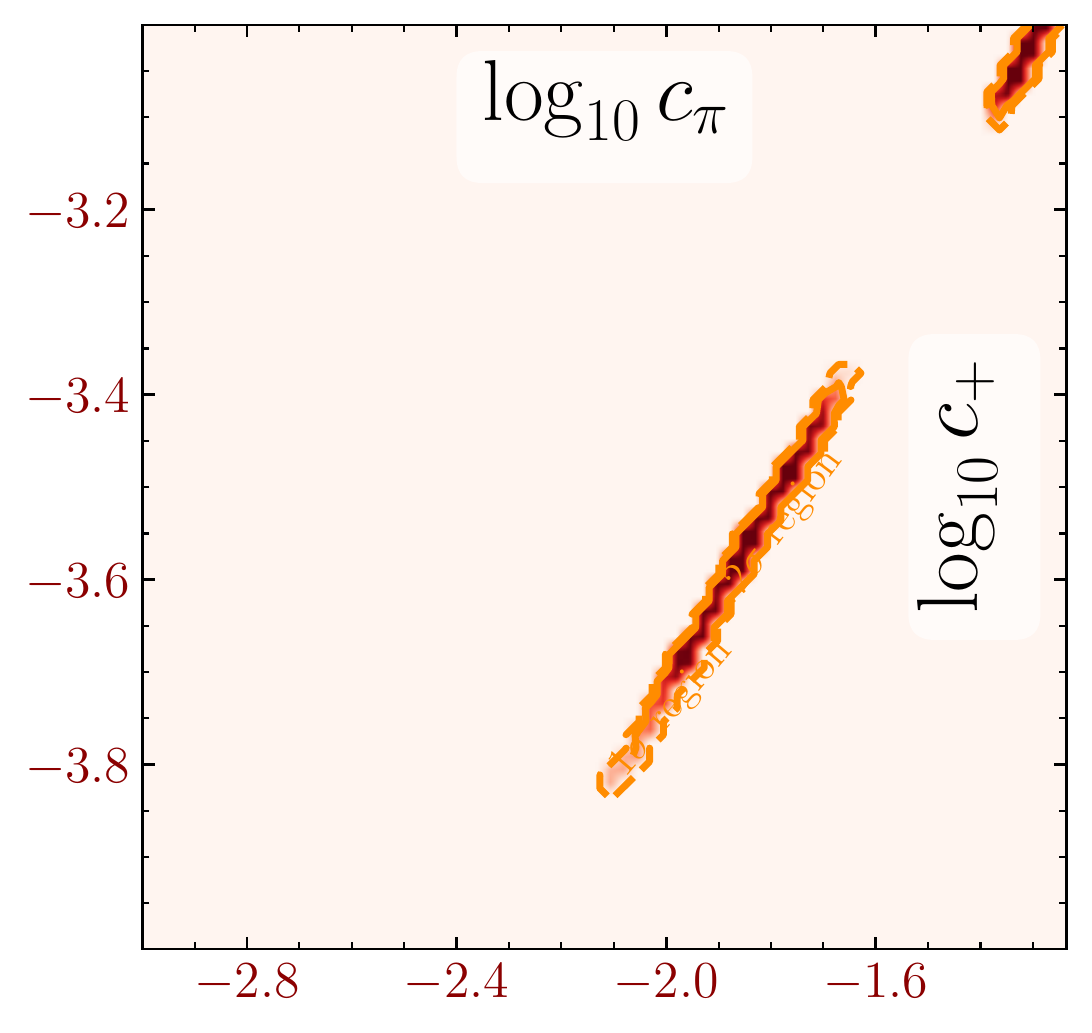}
		\caption[]%
		{{\small Polynomial approximation}}    
		\label{fig:nest-si-poly}
	\end{subfigure}
	\caption[]
	{Two-dimensional profile-likelihood projections of a 3-dimensional
		parameter space with superplot~\citep{Fowlie:2016hew}. Regions of higher likelihood are darker. The data are
		normalized to the maximum observed likelihood.
              }

	\label{fig:nest-si}
\end{figure}

Figure~(\ref{fig:nest-si}) shows
two-dimensional profile-likelihood projections of the
three-dimensional likelihood limits when using the expensive full
simulation, our rational approximation results, and a polynomial
approximation.
Clearly, the rational approximation faithfully represents the full
simulation, while the polynomial approximation is only valid over a
limited range.

\subsection{Tuning HEP Event Generators by Automatic Selection of Observable Weights}

As described earlier, we have developed several algorithms for adjusting the weights used in
our tuning heuristic.
For comparison, we re-consider a tuning exercise performed by the
ATLAS collaboration \cite{ATL-PHYS-PUB-2014-021,Buckley:2014ctn} using a large set of LHC data.

\begin{figure}[ht!]
  \centering
  \includegraphics[width=\textwidth]{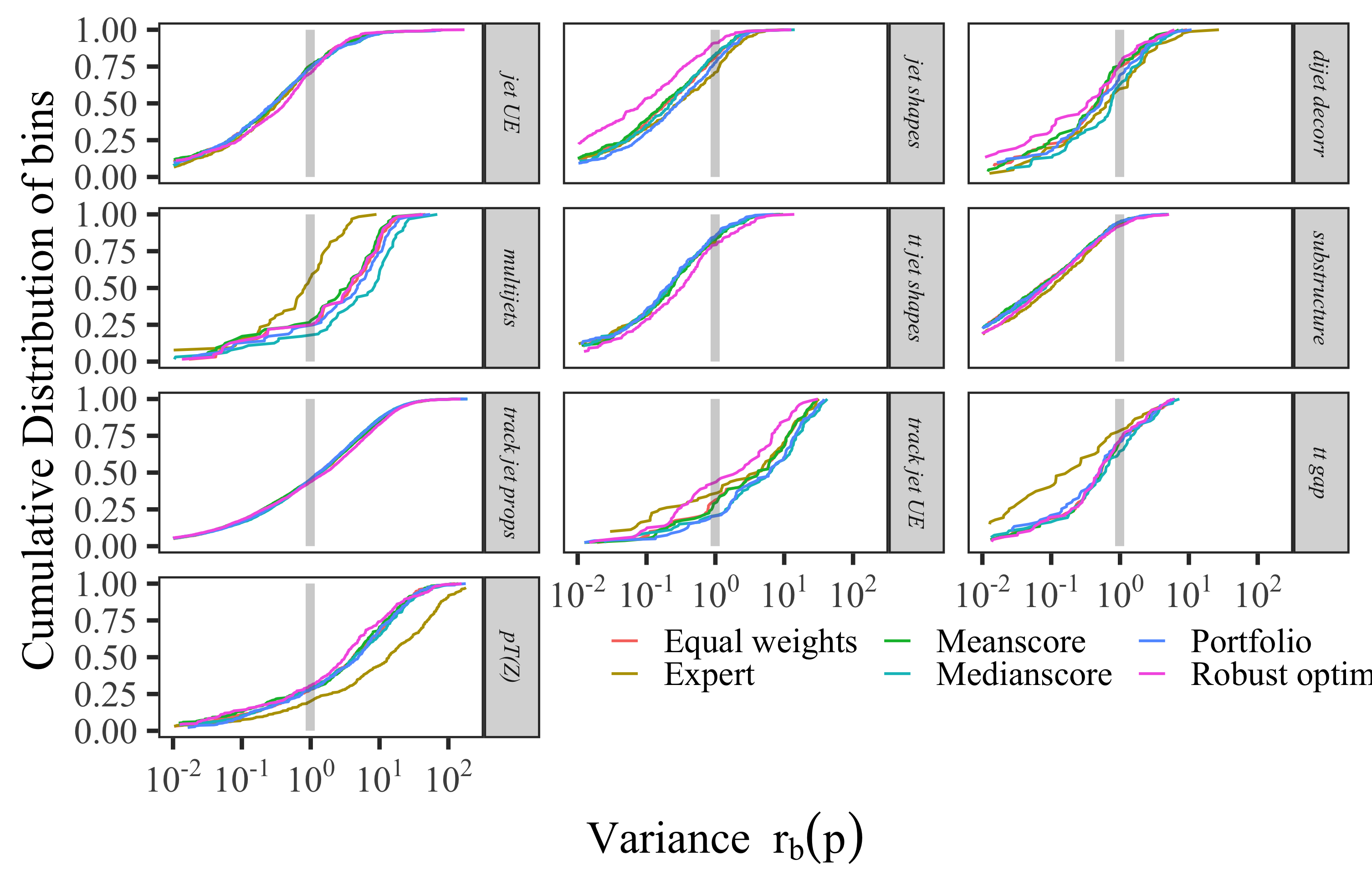}  
  \caption{
    Cumulative distribution of bins (y axis) in each category of the A14 dataset at different bands of variance levels (x axis) given by $r_b(\p)=\frac{\left(f_b(\p) - \calR_b\right)^2}{\Delta f_b(\p)^2+\Delta\calR_b^2}$.
  }
  \label{Fig:chi2perf2binsCat}  
\end{figure}

Figure~\ref{Fig:chi2perf2binsCat} shows cumulative distribution of
$\chi^2$ values per bin for several different classes of observables.
The results of the ATLAS expert tune and a much simpler approach using equal weights
for all observables is also  shown.
Note that the parameters obtained from the robust optimization perform well for \textit{Jet shapes} and \textit{Track-jet UE}. 
We also see that near the variance boundary, the parameters obtained from the \textit{Expert} tune perform better for \textit{Multijets} and \textit{$t \Bar{t}$ gap} whereas the parameters obtained from the other approaches  perform better for \textit{Substructure}. 
These plots demonstrate the trade-off in fitting among the different approaches, which enables the physicist to use these results as guidance for selecting the most appropriate tuning method  depending on the categories that are of greater significance.

%% file: conclusion.tex
\section{Conclusion}

In this paper, we propose an improved software package called \apprentice that enables the construction of pole-free rational approximation, allows for solving the $\chi^2$ minimization problem using a various optimization algorithms that can detect multiple local minima, and provides multiple optimization formulations to automatically assign weights to the observables yielding a less subjective and less time-consuming process to find the optimal event generator tune.
In the future, we are interested in (a) tuning event generators using the MC generator data directly using derivative free optimization methods,  (b) providing non-linear optimization algorithms to minimize $\chi^2$ that are robust against all local minima, and (c) exploring machine learning techniques for dimensionality reduction and for improving the automatic weight assignment toward obtaining optimal generator tune.
If your are interested in these problems or in any other interesting extensions to \apprentice, we look forward to hearing from you.
To contribute, either raise an issue in the \apprentice project or contribute via code by forking the project. The \apprentice project is located at \url{https://github.com/HEPonHPC/apprentice}.